\documentclass[12pt]{article}
\pdfoutput=1

\usepackage{amsmath}
\usepackage{amsfonts}
\usepackage{amscd}
\usepackage{amsthm}
\usepackage{setspace}

\usepackage{graphicx}
\usepackage{authblk}
\usepackage{caption}
\usepackage{ytableau}
\usepackage{mathtools}

\def\Z{\mathbb{Z}}

\def\R{\mathbb{R}}
\def\C{\mathbb{C}}
\def\P{\mathbb{P}}

\def\til{\tilde}

\begin{document}

\begin{titlepage}

\begin{flushright}
KEK-TH-2108
\end{flushright}

\vskip 1cm

\begin{center}

{\bf \Large F-theory models with 3 to 8 U(1) factors on K3 surfaces}

\vskip 1.2cm

Yusuke Kimura$^1$ 
\vskip 0.4cm
{\it $^1$KEK Theory Center, Institute of Particle and Nuclear Studies, KEK, \\ 1-1 Oho, Tsukuba, Ibaraki 305-0801, Japan}
\vskip 0.4cm
E-mail: kimurayu@post.kek.jp

\vskip 2cm
\abstract{\par In this study, we construct four-dimensional F-theory models with 3 to 8 U(1) factors on products of K3 surfaces. We provide explicit Weierstrass equations of elliptic K3 surfaces with Mordell--Weil ranks of 3 to 8. We utilize the method of quadratic base change to glue pairs of rational elliptic surfaces together to yield the aforementioned types of K3 surfaces. The moduli of elliptic K3 surfaces constructed in the study include Kummer surfaces of specific complex structures. We show that the tadpole cancels in F-theory compactifications with flux when these Kummer surfaces are paired with appropriately selected attractive K3 surfaces. We determine the matter spectra on F-theory on the pairs.}  

\end{center}
\end{titlepage}

\tableofcontents
\section{Introduction}
F-theory \cite{Vaf, MV1, MV2} provides an extension of type IIB superstring to a nonperturbative regime and physical information (such as gauge groups and matter representations) are determined via the geometry of the compactification space in F-theory formulation. Compactification space of F-theory has a genus-one fibration, and the modular parameter of a genus-one curve, as a fiber of a genus-one fibration, is identified with the axio-dilaton in F-theory formulation. 
\par 7-branes are wrapped on the components of the codimension one locus in the base space along which fibers degenerate to ``singular fibers'' (such locus is referred to as the {\it discriminant locus}), and the non-Abelian gauge group that forms on the 7-branes is specified via the type of the singular fibers over the discriminant component on which 7-branes are wrapped \cite{MV2, BIKMSV}. 
\par Recently, local model buildings of F-theory \cite{DWmodel, BHV, BHV2, DWGUT} were mainly discussed. It is necessary to examine global aspects of F-theory compactification to address the issues of gravity and early universe. We analyze the geometries of F-theory compactifications from the global perspective in this study.
\par When a genus-one fibration admits a section \footnote{F-theory compactifications on elliptic fibrations admitting a global section have been intensively studied. See, e.g., \cite{MorrisonPark, MPW, BGK, BMPWsection, CKP, BGK1306, CGKP, BMPW2013, CKP1307, CKPS, AL, EKY1410, LSW, CKPT, CGKPS, MP2, MPT1610, BMW2017, CL2017, BMW1706, EKY1712, KimuraMizoguchi, EK1802, Kimura1802, LRW2018, TasilectWeigand, EK1808, MizTani2018, TasilectCL, CMPV1811, TT2019} for studies of such models.}, the sections form a group, and this is referred to as the Mordell--Weil group. The condition that a genus-one fibration has a section geometrically implies that the total space of the genus-one fibration contains a copy of the base space. The Mordell--Weil group is a finitely generated Abelian group, and the rank of the Mordell--Weil group yields the number of U(1) factors in F-theory compactification \cite{MV2}. 
\par F-theory models with various U(1) factors are investigated \cite{MorrisonPark, BMPWsection, CKP, CGKP, BMPW2013, CKPS, MT2014, KMOPR, BGKintfiber, MPT1610, CKPT, Kimura1802, TT2018, CMPV1811, TT2019}. A general F-theory model with one U(1) factor was constructed in \cite{MorrisonPark}. Constructions of F-theory models with two U(1) factors were discussed in \cite{BMPWsection, CKP, CGKP, BMPW2013, CKPT}. Although there are some known examples, the general theory of F-theory Weierstrass models with 3 or more U(1) factors is not fully established. Discussions of F-theory models with 3 or more U(1) factors can be found, for example, in \cite{CKPS, MT2014, BGKintfiber, MPT1610, Kimura1802}. Classification of F-theory models with various U(1) factors, and understanding U(1) gauge groups of high ranks in F-theory model building, are interesting issues to explore. 
\par In this study, we deduce the explicit Weierstrass equations of F-theory models with three to eight U(1) factors. Specifically, we restrict our considerations to elliptically fibered K3 surfaces and construct several families of elliptic K3 surfaces with Mordell--Weil rank 3 to 8. It is generally considerably difficult to directly determine the Mordell--Weil group from the given Weierstrass equation of elliptic K3 surface, and the construction of examples with positive Mordell--Weil ranks is non-trivial. In order to resolve the issue, we utilize the quadratic base change technique (we use ``QBC'' to abbreviate the quadratic base change in this study) used in \cite{Kimura1802} to construct F-theory Weierstrass models with various U(1) factors. 
\par The Picard number of a K3 surface varies over the complex structure moduli, and it is difficult to determine the Mordell--Weil rank of an elliptic K3 surface owing to this property of a K3 surface. An elliptic K3 surface splits into a pair of two 1/2 K3 surfaces in a special limit, which is known as the ``stable degeneration limit'' \cite{FMW, AM} \footnote{F-theory/heterotic duality \cite{Vaf, MV1, MV2, Sen, FMW} is known to strictly hold in this limit. See, e.g., \cite{AHK, BKW, BKL, CGKPS, MizTan, KRES} for recent studies of the stable degeneration in F-theory/heterotic duality.}. The 1/2 K3 surface is also referred to as a rational elliptic surface \footnote{The Mordell-Weil groups of the rational elliptic surfaces with a global section were classified in \cite{OS}.}. In contrast to a K3 surface, the Picard number of a rational elliptic surface does not depend on the complex structure and is constant over the complex structure moduli. Given this property, the sum of the rank of the singularity type and Mordell--Weil rank of a rational elliptic surface with a section is always 8. It is relatively easy to determine types of the singular fibers from the Weierstrass equation, and the Mordell--Weil rank is immediately obtained once the singularity type is computed because the rank of the Mordell--Weil group is 8 minus the rank of the singularity type for a rational elliptic surface. 
\par Gluing a pair of identical rational elliptic surfaces yields an elliptic K3 surface, and QBC describes the operation \cite{KRES}. The QBC can be viewed as the reverse of the stable degeneration in which an elliptic K3 surface splits into a pair of identical rational elliptic surfaces \cite{KRES}. In this study, we apply the QBC method to rational elliptic surfaces with Mordell--Weil ranks 3 to 8 to construct various families of elliptic K3 surfaces with the Mordell--Weil ranks 3 to 8 \footnote{Families of elliptic K3 surfaces of Mordell--Weil ranks 3 and 4 as constructed in this note exhibit singularity types that are different from those of the families of elliptic K3 surface of Mordell--Weil ranks 3 and 4 as constructed in \cite{Kimura1802}.}. F-theory compactifications on the surfaces times a K3 surface yield models with three to eight U(1) factors. Similar constructions of elliptic K3 surfaces with Mordell--Weil ranks 1 to 4 via utilizing the QBC of rational elliptic surfaces can be found in \cite{Kimura1802}.
\par Examples of F-theory models with 3 to 8 U(1) gauge fields as constructed in this study can aid in understanding the general theory of models with multiple U(1) gauge factors. The examples can help in constructing the general theory of F-theory models with multiple U(1)s. 
\par We consider F-theory compactifications on the products of K3 surfaces, and they yield four-dimensional theories with $N=2$ supersymmetry. The enhanced supersymmetry imposes a strong anomaly cancellation condition on the theories. We confirm that the Weierstrass equations deduced in this study are in agreement with the constraint. 
\par We also discuss matter spectra on F-theory on the constructed families of elliptic K3 surfaces times a K3 surface. We find that specific Kummer surfaces belong to the moduli, and the gauge groups forming in F-theory compactifications on the surfaces contain three and four U(1) factors. We determine explicit matter spectra in four-dimensional F-theory compactifications with $N=1$ supersymmetry on the surfaces times a K3 surface with flux \cite{BB, SVW, W, GVW, DRS} \footnote{Recent progress of four-form fluxes in F-theory can be found, e.g., in \cite{BCV, MS, KMW, KMW2, CGK, BKL, CKPOR, SNW, LW}.}.
\par In F-theory, matter representations arise from local rank-one enhancements \cite{BIKMSV, KV, GM, MTmatter, GM2} of the singularities of a compactification space. Other types of matter representations arising from the structure of divisor were discussed, e.g., in \cite{KMP, Pha}. \cite{KM} discussed the deformations and resolutions of singularities.
\par The types of the singular fibers of elliptically fibered surfaces were classified by Kodaira \cite{Kod1, Kod2}, and methods to study the singular fibers of elliptic surfaces were discussed in \cite{Ner, Tate}. The types of the singular fibers of elliptically fibered K3 surfaces, and the corresponding singularity types in F-theory compactifications \cite{MV2, BIKMSV} are presented in Table \ref{tab fibertypeandADEsingularity}.

\begingroup
\renewcommand{\arraystretch}{1.1}
\begin{table}[htb]
\begin{center}
  \begin{tabular}{|c|c|} \hline
Type of singular fiber & $\begin{array}{c}
\mbox{Singularity type} \\
\mbox{of the compactification space}
\end{array}$ \\ \hline
$I_1$, $II$ & none. \\
$I_n$ ($n\ge 2$) & $A_{n-1}$ \\
$I^*_m$ ($m\ge 0$) & $D_{m+4}$ \\
$III$ & $A_1$ \\
$IV$ & $A_2$ \\
$II^*$ & $E_8$ \\
$III^*$ & $E_7$ \\
$IV^*$ & $E_6$ \\ \hline   
\end{tabular}
\caption{Types of singular fibers and the corresponding singularities.}
\label{tab fibertypeandADEsingularity}
\end{center}
\end{table}  
\endgroup

\par This study is structured as follows: in section \ref{sec2}, we briefly review the techniques of the quadratic base change (QBC) utilized in \cite{Kimura1802} to construct F-theory models with various U(1) factors. F-theory on the products of K3 surfaces yield 4d theory with enhanced $N=2$ supersymmetry. Anomaly cancellation condition requires that the total number of the 7-branes is 24, and we can confirm the consistency of the F-theory models obtained in this study by using this. We also review this point. In section \ref{sec3}, we construct families of rational elliptic surfaces with Mordell--Weil ranks 3 to 8. We select specific singularity types, and we deduce the Weierstrass equations of the surfaces. 
\par In section \ref{sec4}, we apply the QBC method to rational elliptic surfaces constructed in section \ref{sec3} to yield elliptic K3 surfaces with Mordell--Weil ranks 3 to 8. We deduce Weierstrass equations of these K3 surfaces. F-theory compactifications on the resulting spaces yield models with 3 to 8 U(1) gauge fields. Similar organization can be found in \cite{Kimura1802}. We discuss matter spectra on F-theory on the spaces times a K3 surface in section \ref{sec5}. We state our concluding remarks in section \ref{sec6}.

\section{Review of quadratic base change and anomaly cancellation condition}
\label{sec2}

\subsection{Review of quadratic base change}
\label{ssec2.1}
We briefly review the method of quadratic base change (QBC) of rational elliptic surfaces to yield elliptically fibered K3 surfaces. 
\par In the stable degeneration limit, an elliptic K3 surface on the F-theory side splits into a pair of rational elliptic surfaces. The Picard number of a rational elliptic surface is 10, and this value does not depend on the complex structure. The Shioda--Tate formula \cite{Shiodamodular, Tate1, Tate2} for elliptic surface $M$ states that the following equality holds:
\begin{equation}
\label{ST formula in 2.1}
\rho(M)=2+{\rm rk} \, ADE + {\rm rk} \, MW(M).
\end{equation}
$ADE$ in (\ref{ST formula in 2.1}) denotes the singularity type of an elliptic surface $M$, and $\rho(M)$ denotes the Picard number of an elliptic surface $M$. When this formula is applied to a rational elliptic surface $X$, because the Picard number of rational elliptic surface $X$ is 10, $\rho(X)=10$, the following equality is obtained:
\begin{equation}
\label{RES sum 8 in 2.1}
{\rm rk} \, ADE + {\rm rk} \, MW(X) = 8.
\end{equation}
This implies that the sum of the rank of the singularity type and the Mordell--Weil rank of a rational elliptic surface is always 8. 
\par When a rational elliptic surface has a global section, the quadratic base change corresponds to an operation that replaces each of the coordinate variables $t$ and $s$ of the coordinate $[t:s]$ of the base $\P^1$ with quadratic polynomials:
\begin{eqnarray}
\label{QBC in 2.1}
t & \rightarrow & \alpha_1 \, t^2+ \alpha_2 \, ts + \alpha_3 \, s^2 \\ \nonumber
s & \rightarrow & \alpha_4 \, t^2+ \alpha_5 \, ts + \alpha_6 \, s^2.
\end{eqnarray}
We used $\alpha_i$, $i=1, \cdots, 6$, to denote the parameters of QBC. The surface that results from applying this operation to a rational elliptic surface is an elliptic K3 surface. When an original rational elliptic surface is given by the Weierstrass equation as follows:
\begin{equation}
y^2=x^3+ f_4(t,s)\, x+ g_6(t,s),
\end{equation}
where $f_4$ and $g_6$ denote homogeneous polynomials of degrees 4 and 6, respectively, in $t,s$, the resulting K3 surface is given by the Weierstrass form as follows:
\begin{equation}
\begin{split}
y^2= & x^3+ f_4(\alpha_1 \, t^2+ \alpha_2 \, ts + \alpha_3 \, s^2, \alpha_4 \, t^2+ \alpha_5 \, ts + \alpha_6 \, s^2)\, x \\
& + g_6(\alpha_1 \, t^2+ \alpha_2 \, ts + \alpha_3 \, s^2, \alpha_4 \, t^2+ \alpha_5 \, ts + \alpha_6 \, s^2).
\end{split}
\end{equation}
\par As discussed in \cite{KRES}, QBC can be seen as the reverse of the stable degeneration in which an elliptic K3 surface splits into a pair of identical rational elliptic surfaces. Thus, QBC is geometrically an operation that glues a pair of identical rational elliptic surfaces to yield an elliptic K3 surface. The singular fibers of the resulting K3 surface are twice the number of those of the original rational elliptic surface \cite{KRES}. Thus, the non-Abelian gauge group formed in F-theory compactification on the resulting K3 surface $S$ exhibits the corresponding $ADE$ type twice the singularity type of the original rational elliptic surface $X$. With respect to the special values of the parameters, singular fibers collide and the singularity type of K3 surface is enhanced as discussed in \cite{KRES}. This corresponds to an enhancement of a non-Abelian gauge group in F-theory compactification. 
\par As explained in \cite{Kimura1802}, a section to an original rational elliptic surface $X$ naturally lifts to a section to a resulting elliptic K3 surface $S$, which is obtained as QBC of the rational elliptic surface $X$. Thus, there is a natural embedding of the Mordell--Weil group $MW(X)$ of the original rational elliptic surface $X$ into the Mordell--Weil group $MW(S)$ of the resulting K3 surface $S$. Therefore, the Mordell--Weil group of the original rational elliptic surface $X$ is a subgroup of that of the resulting K3 surface:
\begin{equation}
MW(X) \subset MW(S).
\end{equation}
Thus, the following inequality holds:
\begin{equation}
{\rm rk}\, MW(X) \le {\rm rk}\, MW(S).
\end{equation}
With respect to a generic quadratic base change (namely, for generic values of the parameters of QBC), the following equality holds:
\begin{equation}
{\rm rk}\, MW(X) = {\rm rk}\, MW(S).
\end{equation}
This is shown in \cite{Kimura1802}. Here, we provide a different proof. (See also \cite{Kloosterman, SchShio}.) We fix the singularity rank of a rational elliptic surface $X$ as $n$. ($0\le n \le 8$.) Then the Mordell--Weil of the rational elliptic surface $X$ corresponds to $8-n$ by the equation (\ref{RES sum 8 in 2.1}):
\begin{equation}
\label{rk MW RES in 2.1}
{\rm rk}\, MW(X) = 8-n.
\end{equation}
Given that $MW(X)$ is a subgroup of $MW(S)$, the K3 surface $S$ has the Mordell--Weil rank at least $8-n$. Subsequently, the elliptic K3 surface $S$ obtained as QBC of the rational elliptic surface $X$ exhibits singularity rank $2n$. Furthermore, the N\'eron-Severi lattice of K3 surface $S$ contains a rank 2 sublattice generated by a zero-section and a smooth fiber. Thus, the following inequality holds for the Picard number of the K3 surface $S$:
\begin{equation}
\label{lower bound 10+n in 2.1}
\rho(S) \ge (8-n)+2n+2=10+n.
\end{equation}
The original rational elliptic surface $X$ has moduli of complex deformations of dimension $8-n$. Additionally, QBC contains two genuine parameters of complex deformations. Therefore, the resulting K3 surface $S$ has the complex moduli of the following dimension: 
\begin{equation}
8-n+2=10-n.
\end{equation}
\par K3 surface of Picard number $\rho$ has the complex structure moduli of dimension $20-\rho$. If a generic member $S$ of K3 surfaces obtained as QBC of the rational elliptic surface $X$ exhibits Mordell--Weil rank strictly exceeding the rank of $MW(X)$, then it has a Picard number strictly exceeding the lower bound $10+n$ in (\ref{lower bound 10+n in 2.1}). If this occurs, generic members $S$ of the family of K3 surfaces obtained as QBC of the family of rational elliptic surfaces $X$ (wherein the singularity type is fixed and its rank is fixed to $n$) have Picard number equal to or greater than $11+n$. Then, the family of K3 surfaces $S$ obtained as the result of QBC has the complex structure moduli of dimension $9-n$ (or smaller), and this contradicts the fact that the family has the complex structure moduli of dimension $10-n$. Thus, we conclude that a generic K3 surface $S$ obtained as the result of QBC has Mordell--Weil rank equal to the Mordell--Weil rank of the original rational elliptic surface $X$.
\par With respect to special values of the parameters, the Mordell--Weil rank of resulting K3 surface enhances \footnote{Examples of enhancements of Mordell--Weil ranks of elliptic K3 surfaces over the moduli of complex structures and appearance of U(1) are discussed in \cite{KimuraMizoguchi}.}.
\par F-theory models with multiple U(1) gauge fields are investigated in recent studies. With respect to F-theory Weierstrass models with 3 U(1) factors or more, the general theory is not fully understood. In this study, we construct examples of families of elliptically fibered K3 surfaces with Mordell--Weil ranks 3 to 8, and we provide explicit Weierstrass equations of the surfaces in section \ref{sec4}. They can aid in understanding the general theory of models with three or more U(1) factors. 
\par First, we construct rational elliptic surfaces with Mordell--Weil ranks 3 to 8 with specific singularity types in section \ref{sec3}. When the singularity types are known, the Mordell--Weil ranks are immediately determined owing to the equation (\ref{RES sum 8 in 2.1}). We utilize the fact that the rank of the Mordell--Weil group is generically invariant under QBC to obtain families of elliptic K3 surfaces with Mordell--Weil ranks 3 to 8 in section \ref{sec4} via applying QBC to rational elliptic surfaces that will be constructed in section \ref{sec3}. 
\par Families of rational elliptic surfaces with specific singularity types with the Mordell--Weil ranks 1 to 4 were constructed in \cite{Kimura1802}. QBC was applied to these surfaces, and Weierstrass equations of elliptic K3 surfaces with Mordell--Weil ranks 1 to 4 were obtained in \cite{Kimura1802}. 

\subsection{Anomaly cancellation condition}
\label{ssec2.2}
F-theory compactifications on an elliptic K3 surface times a K3 surface yields a 4d $N=2$ theory. The enhanced supersymmetry imposes a strong constraint on the theory. The tadpole cancellation condition without flux indicates that the total number of 7-branes present in the compactification should be 24 \cite{K}. We can perform a consistency check of F-theory models using this condition. 
\par The base of elliptic K3 $\times$ K3 is $\P^1 \times$ K3, and the condition to cancel the tadpole without a flux determines the form of the discriminant locus as $\{ 24 \hspace{2mm} {\rm points}\}$ $\times$ K3 \cite{K}. 7-branes are wrapped on K3 surfaces in the base, and they are parallel \cite{K}. Here, 24 points in the form of the discriminant locus are counted with multiplicity assigned.
\par The Euler number of singular fiber \cite{Kod2} yields the number of the 7-branes wrapped on the component over which the singular fibers lie. The numbers of the 7-branes for the types of singular fibers of an elliptic K3 surface are listed in Table \ref{table fiber types and 7-branes in 2.2}. Euler numbers of the singular fibers of elliptic surfaces were computed in \cite{Kod2}.

\begingroup
\renewcommand{\arraystretch}{1.1}
\begin{table}[htb]
\centering
  \begin{tabular}{|c|c|} \hline
fiber type & $
\begin{array}{c}
\mbox{\# of 7-branes} \\
\mbox{(Euler number)} \\
\end{array} $ \\ \hline
$I_n$ & $n$\\
$I^*_m$ &  $m+$6\\ 
$I_0^*$ &  6\\ 
$II$ &  2\\
$III$ & 3\\
$IV$ & 4\\
$IV^*$ & 8\\ 
$III^*$ & 9\\
$II^*$ & 10\\ \hline
\end{tabular}
\caption{\label{table fiber types and 7-branes in 2.2}List of fiber types and associated numbers of 7-branes.}
\end{table}
\endgroup

\par In section \ref{ssec4.2}, we perform the consistency checks of the resulting F-theory compactifications by confirming that this condition is satisfied. 

\section{Rational elliptic surfaces with Mordell--Weil ranks 3 to 8}
\label{sec3}

\subsection{Mordell--Weil rank 8}
\label{ssec3.1}
When the Weierstrass coefficients of a rational elliptic surface are generic, then all the singular fibers of the surface have type $I_1$, and its singularity type is trivial, i.e., it has rank 0. Given (\ref{RES sum 8 in 2.1}), the Mordell--Weil rank of the surface is 8.
\par We may assume that two singular fibers are located at $[t:s]=[1:0], [0:1]$. This amounts to require that the discriminant is divisible by the factor $t\, s$. Given this condition, the Weierstrass equation of the rational elliptic surface is as follows:
\begin{equation}
\label{RES rk 8 in 3.1}
\begin{split}
y^2= &  x^3+ (-3a^2\, t^4+ b\, t^3s+ c\, t^2s^2+d\, ts^3-3\, s^4)\, x \\
& +(-2a^3\, t^6+e\, t^5s+f\, t^4s^2+ g\, t^3s^3+ h\, t^2s^4+j\, ts^5-2\, s^6).
\end{split}
\end{equation}
$a,b,c,d,e,f,g,h,j$ denote parameters, and one of them is superfluous. Three singular fibers can be sent to specific positions under an automorphism of the base $\P^1$, and thus we also assume that another type $I_1$ fiber is located at $[t:s]=[1:1]$. The condition reduces the number of the parameters by one, and thus the actual number of parameters is 8. 
\par Generic members in the family of rational elliptic surfaces (\ref{RES rk 8 in 3.1}) have Mordell--Weil rank 8, and the family of rational elliptic surfaces (\ref{RES rk 8 in 3.1}) exhibits eight-dimensional complex structure moduli. 
\par The discriminant is given as follows:
\begin{equation}
\begin{split}
\Delta = & t\, s\cdot \Big[ (108d-108j)\, s^{10}+(108c-36d^2-108h+27j^2)\, ts^{9} \\
& +(108b-72cd+4d^3-108g+54hj)\, t^2s^8 \\
& +(-324a^2-36c^2-72bd+12cd^2-108f+27h^2+54gj)\, t^3s^7 \\
& +(-72bc+216a^2d+12c^2d+12bd^2-108e+54gh+54fj)\, t^4s^6 \\
& + (216a^3-36b^2+216a^2c+4c^3+24bcd-36a^2d^2+27g^2+54fh+54ej)\, t^5s^5 \\
& + (216a^2b+12bc^2+12b^2d-72a^2cd+54fg+54eh-108a^3j)\, t^6s^4 \\
& + (-324a^4+12b^2c-36a^2c^2-72a^2bd+27f^2+54eg-108a^3h)\, t^7s^3 \\
& + (4b^3-72a^2bc+108a^4d+54ef-108a^3g)\, t^8s^2 \\
& + (-36a^2b^2+108a^4c+27e^2-108a^3f)\, t^9s + (108a^4b-108a^3e)\, t^{10} \Big].
\end{split}
\end{equation}

\par Rational elliptic surface (\ref{RES rk 8 in 3.1}) generically has twelve type $I_1$ fibers. The vanishing orders of the coefficients, $f_4$ and $g_6$, of the Weierstrass equation $y^2=x^3+ f_4 \, x+g_6$, and the corresponding types of the singular fibers are shown in Table \ref{table ordersandtypesoffibers}.

\begingroup
\renewcommand{\arraystretch}{1.5}
\begin{table}[htb]
\begin{center}
  \begin{tabular}{|c|c|c|c|} \hline
$
\begin{array}{c}
\mbox{Type of} \\
\mbox{singular fiber}
\end{array}
$
 & ord($f_4$) & ord($g_6$) & ord($\Delta$) \\ \hline
$I_0 $ & $\ge 0$ & $\ge 0$ & 0 \\ \hline
$I_n $  ($n\ge 1$) & 0 & 0 & $n$ \\ \hline
$II $ & $\ge 1$ & 1 & 2 \\ \hline
$III $ & 1 & $\ge 2$ & 3 \\ \hline
$IV $ & $\ge 2$ & 2 & 4 \\ \hline
$I_m^*$  ($m \ge 1$) & 2 & 3 & $m+6$ \\ \hline
$I_0^*$ & $\ge 2$ & 3 & 6 \\ \cline{2-3}
 & 2 & $\ge 3$ &  \\ \hline
$IV^*$ & $\ge 3$ & 4 & 8 \\ \hline
$III^*$ & 3 & $\ge 5$ & 9 \\ \hline
$II^*$ & $\ge 4$ & 5 & 10 \\ \hline   
\end{tabular}
\caption{\label{table ordersandtypesoffibers}Fiber types and the corresponding vanishing orders of the Weierstrass coefficients, $f_4$ and $g_6$, of $y^2=x^3+f_4\, x+g_6$. The orders of the discriminant $\Delta$ are also shown.}
\end{center}
\end{table}  
\endgroup

\subsection{Mordell--Weil rank 7}
\label{ssec3.2}
We construct a family of rational elliptic surface with the singularity type $A_1$. Generic members of the family have Mordell--Weil rank 7 as given by the equation (\ref{RES sum 8 in 2.1}). It is assumed that a type $I_2$ fiber is located at $[t:s]=[0:1]$. The Weierstrass equation of a member of the family is given as follows:
\begin{equation}
\label{RES rk 7 in 3.2}
y^2=x^3+ (a\, t^4+b\, t^3s+c\, t^2s^2+d\, ts^3-3s^4)\, x+(e\, t^6+f\, t^5s+g\, t^4s^2+h\, t^3s^3+j\, t^2s^4+d\, ts^5-2s^6),
\end{equation}
with the discriminant 
\begin{equation}
\begin{split}
\Delta = & t^2\cdot \Big[ (108c-9d^2-108j)\, s^{10}+(108b-72cd+4d^3-108h+54dj)\, ts^{9} \\
& +(108a-36c^2-72bd+12cd^2-108g+54dh+27j^2)\, t^2s^8 \\
& +(-72bc-72ad+12c^2d+12bd^2-108f+54dg+54hj)\, t^3s^7 \\
& +(-36b^2-72ac+4c^3+24bcd+12ad^2-108e+54df+27h^2+54gj)\, t^4s^6 \\
& + (-72ab+12bc^2+12b^2d+24acd+54de+54gh+54fj)\, t^5s^5 \\
& + (-36a^2+12b^2c+12ac^2+24abd+27g^2+54fh+54ej)\, t^6s^4 \\
& + (4b^3+24abc+12a^2d+54fg+54eh)\, t^7s^3 + (12ab^2+12a^2c+27f^2+54eg)\, t^8s^2 \\
& + (12a^2b+54ef)\, t^9s + (4a^3+27e^2)\, t^{10} \Big].
\end{split}
\end{equation}
$a,b,c,d,e,f,g,h, j$ denote parameters. Among the parameters, two correspond to redundant parameters. Given an argument similar to that stated in section \ref{ssec3.1}, we can send two type $I_1$ fibers to $[t:s]=[1:0], [1:1]$. The condition that type $I_1$ is located at $[t:s]=[1:0]$ is given as follows
\begin{equation}
4a^3+27e^2=0.
\end{equation}
We can replace $(a,e)$ with $(-3\til{a}^2, -2\til{a}^3)$ to satisfy the condition. A rational elliptic surface (\ref{RES rk 7 in 3.2}) has one type $I_2$ fiber and ten type $I_1$ fibers.
\par Family of rational elliptic surfaces (\ref{RES rk 7 in 3.2}) exhibits seven-dimensional complex structure moduli.

\subsection{Mordell--Weil rank 6}
\label{ssec3.3}
We construct a family of rational elliptic surfaces with singularity type $A_2$. By (\ref{RES sum 8 in 2.1}), generic members of this family have the Mordell--Weil rank 6. It is assumed that a type $I_3$ fiber is located at $[t:s]=[0:1]$. The Weierstrass equation of a member of the family is given as follows:
\begin{equation}
\label{RES rk 6 in 3.3}
y^2=x^3+ (a\, t^4+b\, t^3s+c\, t^2s^2+d\, ts^3-3s^4)\, x+\Big(e\, t^6+f\, t^5s+g\, t^4s^2+h\, t^3s^3+(c-\frac{d^2}{12})\, t^2s^4+d\, ts^5-2s^6\Big),
\end{equation}
with the discriminant 
\begin{equation}
\label{disc rk 6 in 3.3}
\begin{split}
\Delta = & t^3\cdot \Big[ (108b-18cd-\frac{d^3}{2}-108h)\, s^{9} \\
& +(108a-9c^2-72bd+\frac{15}{2}cd^2+\frac{3}{16}d^4-108g+54dh)\, ts^8 \\
& +(-72bc-72ad+12c^2d+12bd^2-108f+54dg+54ch-\frac{9}{2}d^2h)\, t^2s^7 \\
& +(-36b^2-72ac+4c^3+24bcd+12ad^2-108e+54df+54cg-\frac{9}{2}d^2g+27h^2)\, t^3s^6 \\
& + (-72ab+12bc^2+12b^2d+24acd+54de+54cf-\frac{9}{2}d^2f+54gh)\, t^4s^5 \\
& + (-36a^2+12b^2c+12ac^2+24abd+54ce-\frac{9}{2}d^2e+27g^2+54fh)\, t^5s^4 \\
& + (4b^3+24abc+12a^2d+54fg+54eh)\, t^6s^3 + (12ab^2+12a^2c+27f^2+54eg)\, t^7s^2 \\
& + (12a^2b+54ef)\, t^8s + (4a^3+27e^2)\, t^{9} \Big].
\end{split}
\end{equation}
$a,b,c,d,e,f,g,h$ denote parameters. Among the parameters, two correspond to redundant parameters. Two type $I_1$ fibers can be sent to $[t:s]=[1:0], [1:1]$ via an automorphism of the base $\P^1$. The conditions reduce the number of the parameters by two, and thus the actual number of the parameters is 6.
\par Family of rational elliptic surfaces (\ref{RES rk 6 in 3.3}) exhibits six-dimensional complex structure moduli. 

\subsection{Mordell--Weil rank 5}
\label{ssec3.4}
We construct a family of rational elliptic surfaces with the singularity type $A_3$. Generic members of the constructed family have Mordell--Weil rank 5. We assume that a type $I_4$ fiber is located at $[t:s]=[0:1]$. The Weierstrass equation of a member of this family is given as follows:
\begin{equation}
\label{RES rk 5 in 3.4}
\begin{split}
y^2= & x^3+ (a\, t^4+b\, t^3s+c\, t^2s^2+d\, ts^3-3s^4)\, x \\
& +\Big(e\, t^6+f\, t^5s+g\, t^4s^2+(b-\frac{cd}{6}-\frac{d^3}{216})\, t^3s^3+(c-\frac{d^2}{12})\, t^2s^4+d\, ts^5-2s^6\Big),
\end{split}
\end{equation}
with the discriminant 
\begin{equation}
\label{disc rk 5 in 3.4}
\begin{split}
\Delta = & t^4\cdot \Big[ (108a-9c^2-18bd-\frac{3}{2}cd^2-\frac{d^4}{16}-108g)\, s^{8} \\
& +(-18bc-72ad+3c^2d+\frac{15}{2}bd^2+\frac{cd^3}{2}+\frac{d^5}{48}-108f+54dg)\, ts^7 \\
& +(-9b^2-72ac+4c^3+15bcd+12ad^2+\frac{3}{4}c^2d^2-\frac{bd^3}{4}+\frac{cd^4}{24}+\frac{d^6}{1728}-108e+54df+54cg-\frac{9}{2}d^2g)\, t^2s^6 \\
& +(-72ab+12bc^2+12b^2d+24acd+54de+54cf-\frac{9}{2}d^2f+54bg-9cdg-\frac{d^3g}{4})\, t^3s^5 \\
& + (-36a^2+12b^2c+12ac^2+24abd+54ce-\frac{9}{2}d^2e+54bf-9cdf-\frac{d^3f}{4}+27g^2)\, t^4s^4 \\
& + (4b^3+24abc+12a^2d+54be-9cde-\frac{d^3e}{4}+54fg)\, t^5s^3 \\
& + (12ab^2+12a^2c+27f^2+54eg)\, t^6s^2 + (12a^2b+54ef)\, t^7s \\
& + (4a^3+27e^2)\, t^8 \Big].
\end{split}
\end{equation}
$a,b,c,d,e,f,g$ denote parameters. Among these parameters, two correspond to redundant parameters. Two type $I_1$ fibers can be sent to $[t:s]=[1:0], [1:1]$ under an automorphism of the base $\P^1$. The conditions reduce the number of the parameters by two, and the actual number of the parameters is 5.
\par Family of rational elliptic surfaces (\ref{RES rk 5 in 3.4}) has five-dimensional complex structure moduli.

\subsection{Mordell--Weil rank 4}
\label{ssec3.5}
We construct three families of rational elliptic surfaces, the singularity types of which are $A_3A_1$, $A_2^2$, and $D_4$. Generic members of the families have Mordell--Weil rank 4.
\par Weierstrass equations of a family of rational elliptic surfaces with singularity type $A_3 A_1$ can be obtained from the equation (\ref{RES rk 5 in 3.4}), via imposing a condition on the equation (\ref{RES rk 5 in 3.4}) that it has a type $I_2$ fiber at the infinity, $[t:s]=[1:0]$. This is equivalent to requiring that the coefficients of the terms $t^{11}s$ and $t^{12}$ vanish in the discriminant (\ref{disc rk 5 in 3.4}). The coefficient of $t^{12}$ is as follows: 
\begin{equation}
4a^3+27e^2,
\end{equation}
and this vanishes when we replace $(a,e)$ with $(-3a^2, 2a^3)$. Following this replacement, the coefficient of the term $t^{11}s$ vanishes when we set the following:
\begin{equation}
f=-ab.
\end{equation}
Thus, we deduce that the following Weierstrass equations describe the family of rational elliptic surfaces with singularity type $A_3A_1$:
\begin{equation}
\label{RES A3A1 in 3.5}
\begin{split}
y^2= & x^3+ (-3a^2\, t^4+b\, t^3s+c\, t^2s^2+d\, ts^3-3s^4)\, x \\
& +\Big(2a^3\, t^6-ab\, t^5s+g\, t^4s^2+(b-\frac{cd}{6}-\frac{d^3}{216})\, t^3s^3+(c-\frac{d^2}{12})\, t^2s^4+d\, ts^5-2s^6\Big),
\end{split}
\end{equation}
$a,b,c,d,g$ denote parameters. We can assume that one of the type $I_1$ fibers is located at $[t:s]=[1:1]$, and this condition reduces the number of parameters by one. Therefore, the actual number of the parameters is 4. 
\par The discriminant is given as follows:
\begin{equation}
\begin{split}
\Delta = & t^4\cdot \Big[ (-324a^2-9c^2-18bd-\frac{3}{2}cd^2-\frac{d^4}{16}-108g)\, s^{8} \\
& +(108ab-18bc+216a^2d+3c^2d+\frac{15}{2}bd^2+\frac{cd^3}{2}+\frac{d^5}{48}+54dg)\, ts^7 \\
& +(-216a^3-9b^2+216a^2c+4c^3-54abd+15bcd\\
& -36a^2d^2+\frac{3}{4}c^2d^2-\frac{bd^3}{4}+\frac{cd^4}{24}+\frac{d^6}{1728}+54cg-\frac{9}{2}d^2g)\, t^2s^6 \\
& +(216a^2b-54abc+12bc^2+108a^3d+12b^2d-72a^2cd+\frac{9}{2}abd^2+54bg-9cdg-\frac{d^3g}{4})\, t^3s^5 \\
& + (-324a^4-54ab^2+108a^3c+12b^2c-36a^2c^2-72a^2bd+9abcd-9a^3d^2+\frac{abd^3}{4}+27g^2)\, t^4s^4 \\
& + (108a^3b+4b^3-72a^2bc+108a^4d-18a^3cd-\frac{a^3d^3}{2}-54abg)\, t^5s^3 \\
& + (-9a^2b^2+108a^4c+108a^3g)\, t^6s^2  \Big].
\end{split}
\end{equation}

\vspace{5mm}

\par The Weierstrass equations of a family of rational elliptic surfaces with the singularity type $A_2^2$ is obtained from the equation (\ref{RES rk 6 in 3.3}) deduced in section \ref{ssec3.3}, via imposing a condition on the equation (\ref{RES rk 6 in 3.3}) wherein it has a type $I_3$ fiber at the infinity, $[t:s]=[1:0]$. This is equivalent to the condition that the coefficients of the terms $t^{10}s^2$, $t^{11}s$, and $t^{12}$ vanish in the discriminant (\ref{disc rk 6 in 3.3}). The coefficient of $t^{12}$ vanishes when we replace $(a,e)$ with $(-3a^2, 2a^3)$ as previously stated. Following this replacement, the coefficient of $t^{11}s$ vanishes when we set $f=-ab$. Then, the Weierstrass equation (\ref{RES rk 6 in 3.3}) becomes as follows:
\begin{equation}
\label{RES A2A2 in 3.5}
\begin{split}
y^2= & x^3+ (-3a^2\, t^4+b\, t^3s+c\, t^2s^2+d\, ts^3-3s^4)\, x \\
& +\Big(2a^3\, t^6-ab\, t^5s+g\, t^4s^2+h\, t^3s^3+(c-\frac{d^2}{12})\, t^2s^4+d\, ts^5-2s^6\Big).
\end{split}
\end{equation}
The coefficient of the term $t^{10}s^2$ vanishes when the following relation is satisfied:
\begin{equation}
\label{condition A2A2 in 3.5}
12ag=b^2-12a^2c.
\end{equation}
$a,b,c,d,h$ denote parameters, and we can assume that one of type $I_1$ fibers is at $[t:s]=[1:1]$. This reduces the number of the parameters by one. 

\vspace{5mm}

\par We construct a family of rational elliptic surfaces with the singularity type $D_4$. It is assumed that a type $I_0^*$ fiber is located at the origin, $[t:s]=[0:1]$. As stated previously, we assume that a type $I_1$ fiber is located at $[t:s]=[1:0]$. Then, the following Weierstrass equation describes the family:
\begin{equation}
\label{RES D4 in 3.5}
y^2=x^3+ t^2\, (-3t^2+a\, ts+b\, s^2)\, x+ t^3\, (2t^3+c\, t^2s+d\, ts^2+e\, s^3).
\end{equation}
$a,b,c,d,e$ denote parameters, and one of the parameters is superfluous. We can also assume that one of the type $I_1$ fibers is located at $[t:s]=[1:1]$, and this condition eliminates the redundancy. 
\par The discriminant is given as follows:
\begin{equation}
\label{disc D4 in 3.5}
\begin{split}
\Delta = & t^6\cdot \Big[ (4b^3+27e^2)\, s^{6} +(12ab^2+54de)\, ts^5 +(12a^2b-36b^2+27d^2+54ce)\, t^2s^4 \\
& +(4a^3-72ab+54cd+108e)\, t^3s^3 + (-36a^2+108b+27c^2+108d)\, t^4s^2 + (108a+108c)\, t^5s  \Big].
\end{split}
\end{equation}

\par The complex structure moduli spaces of rational elliptic surfaces described by the Weierstrass equations (\ref{RES A3A1 in 3.5}), (\ref{RES A2A2 in 3.5}) with the condition (\ref{condition A2A2 in 3.5}) imposed, and (\ref{RES D4 in 3.5}), are four-dimensional. 

\subsection{Mordell--Weil rank 3}
\label{ssec3.6}
We construct two families of rational elliptic surfaces, the singularity types of which are $A_5$ and $D_4A_1$. Generic members of the families have the Mordell--Weil rank 3.
\par We construct a family of rational elliptic surfaces with singularity type $A_5$. We assume that a type $I_6$ fiber is located at the origin, $[t:s]=[0:1]$. The following Weierstrass equation describes this family:
\begin{equation}
\label{RES A5 in 3.6}
\begin{split}
y^2= & x^3+ (a\, t^4+b\, t^3s+c\, t^2s^2+d\,ts^3-3s^4)\, x \\
& +\Big[ e\, t^6+ (-\frac{bc}{6}-\frac{ad}{6}-\frac{c^2d}{72}-\frac{bd^2}{72}-\frac{cd^3}{432}-\frac{d^5}{10368})\, t^5s \\
& +(a-\frac{c^2}{12}-\frac{bd}{6}-\frac{cd^2}{72}-\frac{d^4}{1728})\, t^4s^2 + (b-\frac{cd}{6}-\frac{d^3}{216})\, t^3s^3 + (c-\frac{d^2}{12})\, t^2s^4 + d\, ts^5 -2s^6 \Big].
\end{split}
\end{equation}
Following some computations, we confirm that the Weierstrass equation (\ref{RES A5 in 3.6}) has a type $I_6$ fiber at $[t:s]=[0:1]$ (i.e., the discriminant is divisible by $t^6$), and it also has six type $I_1$ fibers.
\par $a,b,c,d,e$ denote parameters. As stated previously, we can send two type $I_1$ fibers to $[t:s]=[1:0], [1:1]$, and the conditions reduce the number of parameters by two. Therefore, the actual number of the parameters is 3.

\vspace{5mm}

\par We construct a family of rational elliptic surfaces with the singularity type $D_4A_1$. The Weierstrass equations of a family of rational elliptic surfaces with singularity type $D_4A_1$ are obtained from the equation (\ref{RES D4 in 3.5}) as deduced in section \ref{ssec3.5} via imposing a condition on the equation (\ref{RES D4 in 3.5}) wherein it exhibits a type $I_2$ fiber at the infinity, $[t:s]=[1:0]$. The condition is equivalent to the vanishing of the coefficient of the term $t^{11}s$ in the discriminant (\ref{disc D4 in 3.5}). The coefficient of the term $t^{11}s$ is $108(a+c)$, and thus this is satisfied when we set the following expression:
\begin{equation}
c=-a.
\end{equation}
Thus, the following Weierstrass equation yields this family:
\begin{equation}
\label{RES D4A1 in 3.6}
y^2=x^3+ t^2\, (-3t^2+a\, ts+b\, s^2)\, x + t^3\, (2t^3-a\, t^2s+d\, ts^2+e\, s^3),
\end{equation}
with the discriminant 
\begin{equation}
\label{disc D4A1 in 3.6}
\begin{split}
\Delta = & t^6s^2\cdot \Big[ (4b^3+27e^2)\, s^4+(12ab^2+54de)\, ts^3+ (12a^2b-36b^2+27d^2-54ae)\, t^2s^2 \\
& + (4a^3-72ab-54ad+108e)\, t^3s +(-9a^2+108b+108d)\, t^4 \Big].
\end{split}
\end{equation}
From the equations (\ref{RES D4A1 in 3.6}) and (\ref{disc D4A1 in 3.6}), we find that a generic rational elliptic surface (\ref{RES D4A1 in 3.6}) has one type $I^*_0$ fiber at the origin, one type $I_2$ fiber at the infinity, and four type $I_1$ fibers.
\par $a,b,d,e$ denote parameters, and we can send one type $I_1$ fiber such that it lies over the position $[t:s]=[1:1]$. The condition decreases the number of the parameters by one, and thus the actual number of parameters is 3.

\vspace{5mm}

\par The complex structure moduli spaces of rational elliptic surfaces described by the Weierstrass equations (\ref{RES A5 in 3.6}) and (\ref{RES D4A1 in 3.6}) are three-dimensional. 

\section{F-theory models with 3 to 8 U(1) factors}
\label{sec4}

\par In this section, we construct families of elliptically fibered K3 surfaces with Mordell--Weil ranks 3 to 8 via applying QBC to families of rational elliptic surfaces constructed in section \ref{sec3}. Specifically, F-theory compactifications on the resulting K3 surfaces times a K3 surface yield models with 3 to 8 U(1) factors. 

\subsection{Summary}
\label{ssec4.1}
We provide four-dimensional F-theory models with 3 to 8 U(1) gauge fields. The QBC of the families of rational elliptic surfaces constructed in section \ref{sec3} yields elliptic K3 surfaces, the singularity types of which are twice those of the original rational elliptic surfaces. The Mordell--Weil rank of the resulting K3 surface is identical to the Mordell--Weil rank of the original rational elliptic surfaces for generic values of the parameters of QBC. 
\par The construction results in explicit Weierstrass equations of elliptic K3 surfaces with Mordell--Weil ranks 3 to 8. F-theory compactifications on the K3 surfaces times a K3 surface yield models with 3 to 8 U(1) factors. 
\par The F-theory compactifications give 4d theory with $N=2$ supersymmetry. We verify the consistency of the obtained theories by using the anomaly cancellation condition. 

\subsection{F-theory models with multiple U(1)s and explicit Weierstrass equations}
\label{ssec4.2}
\subsubsection{Model with 8 U(1)s}
\label{sssec4.2.1}
QBC replaces the coordinate variables $t,s$ of the base $\P^1$ with quadratic polynomials in the variables $t,s$ as follows:
\begin{eqnarray}
t & \rightarrow & \alpha_1 \, t^2+ \alpha_2 \, ts + \alpha_3 \, s^2 \\ \nonumber
s & \rightarrow & \alpha_4 \, t^2+ \alpha_5 \, ts + \alpha_6 \, s^2,
\end{eqnarray}
as reviewed in section \ref{ssec2.1}. When this is applied to the equation (\ref{RES rk 8 in 3.1}) of rational elliptic surfaces of the Mordell--Weil rank 8 deduced in section \ref{ssec3.1}, the resulting Weierstrass equation is as follows:
\begin{equation}
\label{K3 rk 8 in 4.2.1}
\begin{split}
y^2= &  x^3+ \Big(-3a^2\, (\alpha_1 \, t^2+ \alpha_2 \, ts + \alpha_3 \, s^2)^4+ b\, (\alpha_1 \, t^2+ \alpha_2 \, ts + \alpha_3 \, s^2)^3(\alpha_4 \, t^2+ \alpha_5 \, ts + \alpha_6 \, s^2)\\
& + c\, (\alpha_1 \, t^2+ \alpha_2 \, ts + \alpha_3 \, s^2)^2(\alpha_4 \, t^2+ \alpha_5 \, ts + \alpha_6 \, s^2)^2+d\, (\alpha_1 \, t^2+ \alpha_2 \, ts + \alpha_3 \, s^2)(\alpha_4 \, t^2+ \alpha_5 \, ts + \alpha_6 \, s^2)^3\\
& -3\, (\alpha_4 \, t^2+ \alpha_5 \, ts + \alpha_6 \, s^2)^4 \Big)\, x \\
& +\Big(-2a^3\, (\alpha_1 \, t^2+ \alpha_2 \, ts + \alpha_3 \, s^2)^6+e\, (\alpha_1 \, t^2+ \alpha_2 \, ts + \alpha_3 \, s^2)^5(\alpha_4 \, t^2+ \alpha_5 \, ts + \alpha_6 \, s^2)\\
& +f\, (\alpha_1 \, t^2+ \alpha_2 \, ts + \alpha_3 \, s^2)^4(\alpha_4 \, t^2+ \alpha_5 \, ts + \alpha_6 \, s^2)^2+ g\, (\alpha_1 \, t^2+ \alpha_2 \, ts + \alpha_3 \, s^2)^3(\alpha_4 \, t^2+ \alpha_5 \, ts + \alpha_6 \, s^2)^3\\
& + h\, (\alpha_1 \, t^2+ \alpha_2 \, ts + \alpha_3 \, s^2)^2(\alpha_4 \, t^2+ \alpha_5 \, ts + \alpha_6 \, s^2)^4+j\, (\alpha_1 \, t^2+ \alpha_2 \, ts + \alpha_3 \, s^2)(\alpha_4 \, t^2+ \alpha_5 \, ts + \alpha_6 \, s^2)^5\\
& -2\, (\alpha_4 \, t^2+ \alpha_5 \, ts + \alpha_6 \, s^2)^6 \Big).
\end{split}
\end{equation}
This yields a family of elliptic K3 surfaces with the Mordell--Weil rank 8 wherein the singularity type is generically trivial. Specifically, $a,b,c,d,e,f,g,h,j$ are parameters \footnote{As stated in section \ref{ssec3.1}, one of the parameters is superfluous, and can be expressed in terms of the other parameters.}, and $\alpha_1, \cdots, \alpha_6$ denote the parameters of QBC.
\par Generic member of the family has 24 type $I_1$ fibers. Thus, there are 24 7-branes in F-theory on a K3 surface belonging to the family (\ref{K3 rk 8 in 4.2.1}) times a K3 surface based on Table \ref{table fiber types and 7-branes in 2.2}. This ensures that the anomaly cancellation condition is satisfied. 
\par F-theory compactification on an elliptic K3 surface (\ref{K3 rk 8 in 4.2.1}) times a K3 surface generically does not have a non-Abelian gauge factor. Thus, the gauge group formed in the compactification is as follows: 
\begin{equation}
U(1)^8.
\end{equation}

\subsubsection{Model with 7 U(1)s}
\label{sssec4.2.2}
\par We apply QBC to the family of rational elliptic surfaces (\ref{RES rk 7 in 3.2}) constructed in section \ref{ssec3.2} to yield the following Weierstrass equation:
\begin{equation}
\label{K3 rk 7 in 4.2.2}
\begin{split}
y^2= &  x^3+ \Big(a\, (\alpha_1 \, t^2+ \alpha_2 \, ts + \alpha_3 \, s^2)^4+ b\, (\alpha_1 \, t^2+ \alpha_2 \, ts + \alpha_3 \, s^2)^3(\alpha_4 \, t^2+ \alpha_5 \, ts + \alpha_6 \, s^2)\\
& + c\, (\alpha_1 \, t^2+ \alpha_2 \, ts + \alpha_3 \, s^2)^2(\alpha_4 \, t^2+ \alpha_5 \, ts + \alpha_6 \, s^2)^2+d\, (\alpha_1 \, t^2+ \alpha_2 \, ts + \alpha_3 \, s^2)(\alpha_4 \, t^2+ \alpha_5 \, ts + \alpha_6 \, s^2)^3\\
& -3\, (\alpha_4 \, t^2+ \alpha_5 \, ts + \alpha_6 \, s^2)^4 \Big)\, x \\
& +\Big(e\, (\alpha_1 \, t^2+ \alpha_2 \, ts + \alpha_3 \, s^2)^6+f\, (\alpha_1 \, t^2+ \alpha_2 \, ts + \alpha_3 \, s^2)^5(\alpha_4 \, t^2+ \alpha_5 \, ts + \alpha_6 \, s^2)\\
& +g\, (\alpha_1 \, t^2+ \alpha_2 \, ts + \alpha_3 \, s^2)^4(\alpha_4 \, t^2+ \alpha_5 \, ts + \alpha_6 \, s^2)^2+ h\, (\alpha_1 \, t^2+ \alpha_2 \, ts + \alpha_3 \, s^2)^3(\alpha_4 \, t^2+ \alpha_5 \, ts + \alpha_6 \, s^2)^3\\
& + j\, (\alpha_1 \, t^2+ \alpha_2 \, ts + \alpha_3 \, s^2)^2(\alpha_4 \, t^2+ \alpha_5 \, ts + \alpha_6 \, s^2)^4+d\, (\alpha_1 \, t^2+ \alpha_2 \, ts + \alpha_3 \, s^2)(\alpha_4 \, t^2+ \alpha_5 \, ts + \alpha_6 \, s^2)^5\\
& -2\, (\alpha_4 \, t^2+ \alpha_5 \, ts + \alpha_6 \, s^2)^6 \Big).
\end{split}
\end{equation}
This describes a family of elliptic K3 surfaces with Mordell--Weil rank 7 wherein the singularity type is $A_1^2$. 
\par A generic elliptic K3 surface belonging to the family has two $I_2$ fibers and twenty type $I_1$ fibers. Therefore, there are 24 7-branes in F-theory on K3 surface belonging to this family times a K3 surface. This confirms that the anomaly cancellation condition is satisfied. The anomaly cancellation condition is verified in a similar fashion for the remaining F-theory models.
\par According to Table 1 in \cite{Shimada}, because K3 surface has $A_1^2$ singularity, the torsion part of the Mordell--Weil group of K3 surface (\ref{K3 rk 7 in 4.2.2}) is trivial. Thus, the Mordell--Weil group is 
\begin{equation}
MW \cong \Z^7.
\end{equation}
Therefore, the global structure of the gauge group forming in F-theory compactification on K3 surface (\ref{K3 rk 7 in 4.2.2}) times a K3 surface is 
\begin{equation}
SU(2)^2\times U(1)^7.
\end{equation}

\subsubsection{Model with 6 U(1)s}
\label{sssec4.2.3}
\par We apply QBC to the family of rational elliptic surfaces (\ref{RES rk 6 in 3.3}) constructed in section \ref{ssec3.3} to obtain the following Weierstrass equation:
\begin{equation}
\label{K3 rk 6 in 4.2.3}
\begin{split}
y^2= &  x^3+ \Big(a\, (\alpha_1 \, t^2+ \alpha_2 \, ts + \alpha_3 \, s^2)^4+ b\, (\alpha_1 \, t^2+ \alpha_2 \, ts + \alpha_3 \, s^2)^3(\alpha_4 \, t^2+ \alpha_5 \, ts + \alpha_6 \, s^2)\\
& + c\, (\alpha_1 \, t^2+ \alpha_2 \, ts + \alpha_3 \, s^2)^2(\alpha_4 \, t^2+ \alpha_5 \, ts + \alpha_6 \, s^2)^2+d\, (\alpha_1 \, t^2+ \alpha_2 \, ts + \alpha_3 \, s^2)(\alpha_4 \, t^2+ \alpha_5 \, ts + \alpha_6 \, s^2)^3\\
& -3\, (\alpha_4 \, t^2+ \alpha_5 \, ts + \alpha_6 \, s^2)^4 \Big)\, x \\
& +\Big(e\, (\alpha_1 \, t^2+ \alpha_2 \, ts + \alpha_3 \, s^2)^6+f\, (\alpha_1 \, t^2+ \alpha_2 \, ts + \alpha_3 \, s^2)^5(\alpha_4 \, t^2+ \alpha_5 \, ts + \alpha_6 \, s^2)\\
& +g\, (\alpha_1 \, t^2+ \alpha_2 \, ts + \alpha_3 \, s^2)^4(\alpha_4 \, t^2+ \alpha_5 \, ts + \alpha_6 \, s^2)^2+ h\, (\alpha_1 \, t^2+ \alpha_2 \, ts + \alpha_3 \, s^2)^3(\alpha_4 \, t^2+ \alpha_5 \, ts + \alpha_6 \, s^2)^3\\
& + (c-\frac{d^2}{12})\, (\alpha_1 \, t^2+ \alpha_2 \, ts + \alpha_3 \, s^2)^2(\alpha_4 \, t^2+ \alpha_5 \, ts + \alpha_6 \, s^2)^4\\
& +d\, (\alpha_1 \, t^2+ \alpha_2 \, ts + \alpha_3 \, s^2)(\alpha_4 \, t^2+ \alpha_5 \, ts + \alpha_6 \, s^2)^5  -2\, (\alpha_4 \, t^2+ \alpha_5 \, ts + \alpha_6 \, s^2)^6 \Big).
\end{split}
\end{equation}
This describes a family of elliptic K3 surfaces with the Mordell--Weil rank 6, with singularity type $A_2^2$. 
\par According to Table 1 in \cite{Shimada}, because K3 surface has $A_2^2$ singularity, the torsion part of the Mordell--Weil group of K3 surface (\ref{K3 rk 6 in 4.2.3}) is trivial. Thus, the Mordell--Weil group is 
\begin{equation}
MW \cong \Z^6.
\end{equation}
It can be confirmed that the Mordell--Weil groups of generic members of the remaining families of elliptic K3 surfaces in section \ref{ssec4.2} do not have the torsion part in a similar fashion, therefore we do not repeat this argument for the remaining families of K3 surfaces. The global structure of the gauge group forming in F-theory compactification on K3 surface (\ref{K3 rk 6 in 4.2.3}) times a K3 surface is 
\begin{equation}
SU(3)^2\times U(1)^6.
\end{equation}

\subsubsection{Model with 5 U(1)s}
\label{sssec4.2.4}
\par We apply QBC to the family of rational elliptic surfaces (\ref{RES rk 5 in 3.4}) constructed in section \ref{ssec3.4} to obtain the following Weierstrass equation:
\begin{equation}
\label{K3 rk 5 in 4.2.4}
\begin{split}
y^2= &  x^3+ \Big(a\, (\alpha_1 \, t^2+ \alpha_2 \, ts + \alpha_3 \, s^2)^4+ b\, (\alpha_1 \, t^2+ \alpha_2 \, ts + \alpha_3 \, s^2)^3(\alpha_4 \, t^2+ \alpha_5 \, ts + \alpha_6 \, s^2)\\
& + c\, (\alpha_1 \, t^2+ \alpha_2 \, ts + \alpha_3 \, s^2)^2(\alpha_4 \, t^2+ \alpha_5 \, ts + \alpha_6 \, s^2)^2+d\, (\alpha_1 \, t^2+ \alpha_2 \, ts + \alpha_3 \, s^2)(\alpha_4 \, t^2+ \alpha_5 \, ts + \alpha_6 \, s^2)^3\\
& -3\, (\alpha_4 \, t^2+ \alpha_5 \, ts + \alpha_6 \, s^2)^4 \Big)\, x \\
& +\Big(e\, (\alpha_1 \, t^2+ \alpha_2 \, ts + \alpha_3 \, s^2)^6+f\, (\alpha_1 \, t^2+ \alpha_2 \, ts + \alpha_3 \, s^2)^5(\alpha_4 \, t^2+ \alpha_5 \, ts + \alpha_6 \, s^2)\\
& +g\, (\alpha_1 \, t^2+ \alpha_2 \, ts + \alpha_3 \, s^2)^4(\alpha_4 \, t^2+ \alpha_5 \, ts + \alpha_6 \, s^2)^2 \\
& + (b-\frac{cd}{6}-\frac{d^3}{216})\, (\alpha_1 \, t^2+ \alpha_2 \, ts + \alpha_3 \, s^2)^3(\alpha_4 \, t^2+ \alpha_5 \, ts + \alpha_6 \, s^2)^3\\
& + (c-\frac{d^2}{12})\, (\alpha_1 \, t^2+ \alpha_2 \, ts + \alpha_3 \, s^2)^2(\alpha_4 \, t^2+ \alpha_5 \, ts + \alpha_6 \, s^2)^4\\
& +d\, (\alpha_1 \, t^2+ \alpha_2 \, ts + \alpha_3 \, s^2)(\alpha_4 \, t^2+ \alpha_5 \, ts + \alpha_6 \, s^2)^5  -2\, (\alpha_4 \, t^2+ \alpha_5 \, ts + \alpha_6 \, s^2)^6 \Big).
\end{split}
\end{equation}
This describes a family of elliptic K3 surfaces with the Mordell--Weil rank 5, with the singularity type $A_3^2$. 
\par The Mordell--Weil group of an elliptic K3 surface (\ref{K3 rk 5 in 4.2.4}) is: 
\begin{equation}
MW \cong \Z^5.
\end{equation}
The global structure of the gauge group forming in F-theory compactification on K3 surface (\ref{K3 rk 5 in 4.2.4}) times a K3 surface is 
\begin{equation}
SU(4)^2\times U(1)^5.
\end{equation}

\subsubsection{Model with 4 U(1)s}
\label{sssec4.2.5}
\par We apply QBC to the three families of rational elliptic surfaces (\ref{RES A3A1 in 3.5}), (\ref{RES A2A2 in 3.5}) and (\ref{RES D4 in 3.5}), constructed in section \ref{ssec3.5}. First, we apply QBC to the family of rational elliptic surfaces (\ref{RES A3A1 in 3.5}) with $A_3A_1$ singularity. Consequently, we obtain the following Weierstrass equation:
\begin{equation}
\label{K3 rk 4 A3A1 in 4.2.5}
\begin{split}
y^2= &  x^3+ \Big(-3a^2\, (\alpha_1 \, t^2+ \alpha_2 \, ts + \alpha_3 \, s^2)^4+ b\, (\alpha_1 \, t^2+ \alpha_2 \, ts + \alpha_3 \, s^2)^3(\alpha_4 \, t^2+ \alpha_5 \, ts + \alpha_6 \, s^2)\\
& + c\, (\alpha_1 \, t^2+ \alpha_2 \, ts + \alpha_3 \, s^2)^2(\alpha_4 \, t^2+ \alpha_5 \, ts + \alpha_6 \, s^2)^2+d\, (\alpha_1 \, t^2+ \alpha_2 \, ts + \alpha_3 \, s^2)(\alpha_4 \, t^2+ \alpha_5 \, ts + \alpha_6 \, s^2)^3\\
& -3\, (\alpha_4 \, t^2+ \alpha_5 \, ts + \alpha_6 \, s^2)^4 \Big)\, x \\
& +\Big(2a^3\, (\alpha_1 \, t^2+ \alpha_2 \, ts + \alpha_3 \, s^2)^6-ab\, (\alpha_1 \, t^2+ \alpha_2 \, ts + \alpha_3 \, s^2)^5(\alpha_4 \, t^2+ \alpha_5 \, ts + \alpha_6 \, s^2)\\
& +g\, (\alpha_1 \, t^2+ \alpha_2 \, ts + \alpha_3 \, s^2)^4(\alpha_4 \, t^2+ \alpha_5 \, ts + \alpha_6 \, s^2)^2 \\
& + (b-\frac{cd}{6}-\frac{d^3}{216})\, (\alpha_1 \, t^2+ \alpha_2 \, ts + \alpha_3 \, s^2)^3(\alpha_4 \, t^2+ \alpha_5 \, ts + \alpha_6 \, s^2)^3\\
& + (c-\frac{d^2}{12})\, (\alpha_1 \, t^2+ \alpha_2 \, ts + \alpha_3 \, s^2)^2(\alpha_4 \, t^2+ \alpha_5 \, ts + \alpha_6 \, s^2)^4\\
& +d\, (\alpha_1 \, t^2+ \alpha_2 \, ts + \alpha_3 \, s^2)(\alpha_4 \, t^2+ \alpha_5 \, ts + \alpha_6 \, s^2)^5  -2\, (\alpha_4 \, t^2+ \alpha_5 \, ts + \alpha_6 \, s^2)^6 \Big).
\end{split}
\end{equation}
This equation describes a family of elliptic K3 surfaces with the Mordell--Weil rank 4, with the singularity type $A_3^2 A_1^2$.

\vspace{5mm}

\par Next, we apply QBC to another family of rational elliptic surfaces (\ref{RES D4 in 3.5}) with $D_4$ singularity. This yields the following Weierstrass equation:
\begin{equation}
\label{K3 rk 4 D4 in 4.2.5}
\begin{split}
y^2= & x^3+ (\alpha_1 \, t^2+ \alpha_2 \, ts + \alpha_3 \, s^2)^2\, \Big(-3(\alpha_1 \, t^2+ \alpha_2 \, ts + \alpha_3 \, s^2)^2 \\
& +a\, (\alpha_1 \, t^2+ \alpha_2 \, ts + \alpha_3 \, s^2)(\alpha_4 \, t^2+ \alpha_5 \, ts + \alpha_6 \, s^2)+b\, (\alpha_4 \, t^2+ \alpha_5 \, ts + \alpha_6 \, s^2)^2\Big)\, x \\
& + (\alpha_1 \, t^2+ \alpha_2 \, ts + \alpha_3 \, s^2)^3\, \Big(2(\alpha_1 \, t^2+ \alpha_2 \, ts + \alpha_3 \, s^2)^3 \\
& +c\, (\alpha_1 \, t^2+ \alpha_2 \, ts + \alpha_3 \, s^2)^2(\alpha_4 \, t^2+ \alpha_5 \, ts + \alpha_6 \, s^2) \\
& +d\, (\alpha_1 \, t^2+ \alpha_2 \, ts + \alpha_3 \, s^2)(\alpha_4 \, t^2+ \alpha_5 \, ts + \alpha_6 \, s^2)^2+e\, (\alpha_4 \, t^2+ \alpha_5 \, ts + \alpha_6 \, s^2)^3 \Big).
\end{split}
\end{equation}
This equation describes a family of elliptic K3 surfaces with the Mordell--Weil rank 4, with the singularity type $D_4^2$.

\vspace{5mm}

\par We apply QBC to family of rational elliptic surfaces (\ref{RES A2A2 in 3.5}) with $A_2^2$ singularity to obtain the following Weierstrass form:
\begin{equation}
\label{K3 rk 4 A2A2 in 4.2.5}
\begin{split}
y^2= &  x^3+ \Big(-3a^2\, (\alpha_1 \, t^2+ \alpha_2 \, ts + \alpha_3 \, s^2)^4+ b\, (\alpha_1 \, t^2+ \alpha_2 \, ts + \alpha_3 \, s^2)^3(\alpha_4 \, t^2+ \alpha_5 \, ts + \alpha_6 \, s^2)\\
& + c\, (\alpha_1 \, t^2+ \alpha_2 \, ts + \alpha_3 \, s^2)^2(\alpha_4 \, t^2+ \alpha_5 \, ts + \alpha_6 \, s^2)^2+d\, (\alpha_1 \, t^2+ \alpha_2 \, ts + \alpha_3 \, s^2)(\alpha_4 \, t^2+ \alpha_5 \, ts + \alpha_6 \, s^2)^3\\
& -3\, (\alpha_4 \, t^2+ \alpha_5 \, ts + \alpha_6 \, s^2)^4 \Big)\, x \\
& +\Big(2a^3\, (\alpha_1 \, t^2+ \alpha_2 \, ts + \alpha_3 \, s^2)^6-ab\, (\alpha_1 \, t^2+ \alpha_2 \, ts + \alpha_3 \, s^2)^5(\alpha_4 \, t^2+ \alpha_5 \, ts + \alpha_6 \, s^2)\\
& +g\, (\alpha_1 \, t^2+ \alpha_2 \, ts + \alpha_3 \, s^2)^4(\alpha_4 \, t^2+ \alpha_5 \, ts + \alpha_6 \, s^2)^2 \\
& + h\, (\alpha_1 \, t^2+ \alpha_2 \, ts + \alpha_3 \, s^2)^3(\alpha_4 \, t^2+ \alpha_5 \, ts + \alpha_6 \, s^2)^3\\
& + (c-\frac{d^2}{12})\, (\alpha_1 \, t^2+ \alpha_2 \, ts + \alpha_3 \, s^2)^2(\alpha_4 \, t^2+ \alpha_5 \, ts + \alpha_6 \, s^2)^4\\
& +d\, (\alpha_1 \, t^2+ \alpha_2 \, ts + \alpha_3 \, s^2)(\alpha_4 \, t^2+ \alpha_5 \, ts + \alpha_6 \, s^2)^5  -2\, (\alpha_4 \, t^2+ \alpha_5 \, ts + \alpha_6 \, s^2)^6 \Big).
\end{split}
\end{equation}
The parameters $a,b,c,g$ are subject to the condition:
\begin{equation}
12ag=b^2-12a^2c,
\end{equation}
as stated in section \ref{ssec3.5}. This Weierstrass equation describes a family of elliptic K3 surfaces with the Mordell--Weil rank 4, with the singularity type $A_2^4$.

\par The Mordell--Weil group of the resulting elliptic K3 surfaces (\ref{K3 rk 4 A3A1 in 4.2.5}), (\ref{K3 rk 4 D4 in 4.2.5}) and (\ref{K3 rk 4 A2A2 in 4.2.5}) do not have the torsion part, and they are isomorphic to $\Z^4$: 
\begin{equation}
MW \cong \Z^4.
\end{equation}
\par The global structure of the gauge group forming in F-theory compactification on K3 surface (\ref{K3 rk 4 A3A1 in 4.2.5}) times a K3 surface is 
\begin{equation}
SU(4)^2 \times SU(2)^2 \times U(1)^4.
\end{equation}
\par The global structure of the gauge group in F-theory compactification on K3 surface (\ref{K3 rk 4 D4 in 4.2.5}) times a K3 surface is 
\begin{equation}
SO(8)^2 \times U(1)^4.
\end{equation}
\par The global structure of the gauge group in F-theory compactification on K3 surface (\ref{K3 rk 4 A2A2 in 4.2.5}) times a K3 surface is 
\begin{equation}
SU(3)^4 \times U(1)^4.
\end{equation} 

\subsubsection{Model with 3 U(1)s}
\label{sssec4.2.6}
\par We apply QBC to the two families of rational elliptic surfaces (\ref{RES A5 in 3.6}) and (\ref{RES D4A1 in 3.6}) constructed in section \ref{ssec3.6}. First, we apply QBC to the family of rational elliptic surfaces (\ref{RES A5 in 3.6}) with $A_5$ singularity. Hence, we obtain the following Weierstrass equation:
\begin{equation}
\label{K3 rk 3 A5 in 4.2.6}
\begin{split}
y^2= &  x^3+ \Big(a\, (\alpha_1 \, t^2+ \alpha_2 \, ts + \alpha_3 \, s^2)^4+ b\, (\alpha_1 \, t^2+ \alpha_2 \, ts + \alpha_3 \, s^2)^3(\alpha_4 \, t^2+ \alpha_5 \, ts + \alpha_6 \, s^2)\\
& + c\, (\alpha_1 \, t^2+ \alpha_2 \, ts + \alpha_3 \, s^2)^2(\alpha_4 \, t^2+ \alpha_5 \, ts + \alpha_6 \, s^2)^2+d\, (\alpha_1 \, t^2+ \alpha_2 \, ts + \alpha_3 \, s^2)(\alpha_4 \, t^2+ \alpha_5 \, ts + \alpha_6 \, s^2)^3\\
& -3\, (\alpha_4 \, t^2+ \alpha_5 \, ts + \alpha_6 \, s^2)^4 \Big)\, x \\
& +\Big(e\, (\alpha_1 \, t^2+ \alpha_2 \, ts + \alpha_3 \, s^2)^6 \\
& +(-\frac{bc}{6}-\frac{ad}{6}-\frac{c^2d}{72}-\frac{bd^2}{72}-\frac{cd^3}{432}-\frac{d^5}{10368})\, (\alpha_1 \, t^2+ \alpha_2 \, ts + \alpha_3 \, s^2)^5(\alpha_4 \, t^2+ \alpha_5 \, ts + \alpha_6 \, s^2)\\
& +(a-\frac{c^2}{12}-\frac{bd}{6}-\frac{cd^2}{72}-\frac{d^4}{1728})\, (\alpha_1 \, t^2+ \alpha_2 \, ts + \alpha_3 \, s^2)^4(\alpha_4 \, t^2+ \alpha_5 \, ts + \alpha_6 \, s^2)^2 \\
& + (b-\frac{cd}{6}-\frac{d^3}{216})\, (\alpha_1 \, t^2+ \alpha_2 \, ts + \alpha_3 \, s^2)^3(\alpha_4 \, t^2+ \alpha_5 \, ts + \alpha_6 \, s^2)^3\\
& + (c-\frac{d^2}{12})\, (\alpha_1 \, t^2+ \alpha_2 \, ts + \alpha_3 \, s^2)^2(\alpha_4 \, t^2+ \alpha_5 \, ts + \alpha_6 \, s^2)^4\\
& +d\, (\alpha_1 \, t^2+ \alpha_2 \, ts + \alpha_3 \, s^2)(\alpha_4 \, t^2+ \alpha_5 \, ts + \alpha_6 \, s^2)^5  -2\, (\alpha_4 \, t^2+ \alpha_5 \, ts + \alpha_6 \, s^2)^6 \Big).
\end{split}
\end{equation}
This describes a family of elliptic K3 surfaces with Mordell--Weil rank 3 with singularity type $A_5^2$.

\vspace{5mm}

\par Next, we apply QBC to the second family of rational elliptic surfaces with $D_4A_1$ singularity (\ref{RES D4A1 in 3.6}) to obtain the following Weierstrass equation:
\begin{equation}
\label{K3 rk 3 D4A1 in 4.2.6}
\begin{split}
y^2= & x^3+ (\alpha_1 \, t^2+ \alpha_2 \, ts + \alpha_3 \, s^2)^2\, \Big(-3(\alpha_1 \, t^2+ \alpha_2 \, ts + \alpha_3 \, s^2)^2 \\
& +a\, (\alpha_1 \, t^2+ \alpha_2 \, ts + \alpha_3 \, s^2)(\alpha_4 \, t^2+ \alpha_5 \, ts + \alpha_6 \, s^2)+b\, (\alpha_4 \, t^2+ \alpha_5 \, ts + \alpha_6 \, s^2)^2\Big)\, x \\
& + (\alpha_1 \, t^2+ \alpha_2 \, ts + \alpha_3 \, s^2)^3\, \Big(2(\alpha_1 \, t^2+ \alpha_2 \, ts + \alpha_3 \, s^2)^3 \\
& -a\, (\alpha_1 \, t^2+ \alpha_2 \, ts + \alpha_3 \, s^2)^2(\alpha_4 \, t^2+ \alpha_5 \, ts + \alpha_6 \, s^2) \\
& +d\, (\alpha_1 \, t^2+ \alpha_2 \, ts + \alpha_3 \, s^2)(\alpha_4 \, t^2+ \alpha_5 \, ts + \alpha_6 \, s^2)^2+e\, (\alpha_4 \, t^2+ \alpha_5 \, ts + \alpha_6 \, s^2)^3 \Big).
\end{split}
\end{equation}
This describes a family of elliptic K3 surfaces with the Mordell--Weil rank 3 with singularity type $D_4^2 A_1^2$.

\par The Mordell--Weil group of the resulting elliptic K3 surfaces (\ref{K3 rk 3 A5 in 4.2.6}) and (\ref{K3 rk 3 D4A1 in 4.2.6}) do not have the torsion part, and they are isomorphic to $\Z^3$: 
\begin{equation}
MW \cong \Z^3.
\end{equation}
\par The global structure of the gauge group forming in F-theory compactification on K3 surface (\ref{K3 rk 3 A5 in 4.2.6}) times a K3 surface is 
\begin{equation}
SU(6)^2\times U(1)^3.
\end{equation}
\par The global structure of the gauge group in F-theory compactification on K3 surface (\ref{K3 rk 3 D4A1 in 4.2.6}) times a K3 surface is 
\begin{equation}
SO(8)^2\times SU(2)^2 \times U(1)^3.
\end{equation}

\section{Matter spectra}
\label{sec5}
We discuss the matter spectra on F-theory on the elliptic K3 surfaces that were constructed in section \ref{sec4} times a K3 surface. As discussed in this section, the families of K3 surfaces constructed in section \ref{ssec4.2} include Kummer surfaces. We show that the tadpole can be cancelled in F-theory on the Kummer surfaces times appropriately selected attractive K3 surfaces with the flux turned on. We determine the matter spectra in the compactifications. We briefly review the cancellation of the tadpole in F-theory on the products of K3 surfaces in section \ref{ssec5.1}. In section \ref{ssec5.2}, we discuss the matter spectra on F-theory on the Kummer surfaces belonging to the moduli as constructed in section \ref{ssec4.2}.

\subsection{Review of tadpole cancellation}
\label{ssec5.1}
\par Complex K3 surfaces with the highest Picard number 20 are referred to as attractive \footnote{We follow the convention of the term utilized in \cite{M}. These types of K3 surfaces are termed as ``singular K3 surfaces'' in mathematics.} K3 surfaces. As will be discussed in section \ref{ssec5.2}, Kummer surfaces of specific complex structures belong to the moduli of elliptic K3 surfaces as constructed in section \ref{ssec4.2}, and these Kummer surfaces turn out to be attractive K3 surfaces\footnote{Generally, Kummer surfaces are not necessarily attractive K3 surfaces.}. We briefly review the classification results of the complex structures of the attractive K3 surfaces, and tadpole cancellation in F-theory on the products of attractive K3 surfaces. A review of these topics can be found in \cite{K}. 
\par The orthogonal complement of the N\'eron-Severi lattice $NS$ inside the K3 lattice $\Lambda_{K3}$ is referred to as the ``transcendental lattice''. The complex structures of the attractive K3 surfaces are uniquely specified by the intersection forms of the transcendental lattices \cite{PS-S, SI}.
\par The transcendental lattices of the attractive K3 surfaces are integral and even symmetric $2 \times 2$ matrices, and their intersection forms can be transformed into the following form under the $GL_2(\Z)$ action:
\begin{equation}
\label{int form attractive in 5.1}
\begin{pmatrix}
2a & b \\
b & 2c \\
\end{pmatrix}.
\end{equation}
Parameters $a,b,c$ are integers, $a,b,c\in\Z$, and they satisfy the following relation:
\begin{equation}
a \ge c \ge b \ge 0.
\end{equation}
Thus, the complex structure moduli of the attractive K3 surfaces is labelled by triplets of three integers, namely $a,b,c$. Thus, we denote the attractive K3 surface (in which the transcendental lattice exhibits the intersection form (\ref{int form attractive in 5.1})) as $S_{[2a \hspace{1mm} b \hspace{1mm} 2c]}$. The Kummer surfaces discussed in section \ref{ssec5.2} are attractive K3 surfaces the transcendental lattices of which exhibit the following intersection matrices
\begin{equation}
\begin{pmatrix}
4 & 2 \\
2 & 4 \\
\end{pmatrix}, \hspace{5mm} \begin{pmatrix}
4 & 0 \\
0 & 4 \\
\end{pmatrix}.
\end{equation}
They are denoted as $S_{[4 \hspace{1mm} 2 \hspace{1mm} 4]}$ and $S_{[4 \hspace{1mm} 0 \hspace{1mm} 4]}$, respectively.
\par The tadpole cancellation condition \cite{VW, SVW} for F-theory on K3 $\times$ K3 in the presence of 4-form flux $G$ is expressed as follows:
\begin{equation}
\frac{1}{2}\, \int G\wedge G + N_3 =24.
\end{equation}
We used $N_3$ to denote the number of 3-branes inserted. The 4-form flux $G$ is subject to the quantization condition \cite{W} as follows:
\begin{equation}
G \in H^4(K3\times K3, \Z).
\end{equation}
\par In \cite{AK}, M-theory compactifications on the pairs of attractive K3 surfaces, $S_1\times S_2$, were considered, and the pairs of the complex structures of the attractive K3 surfaces for which the tadpole can be cancelled with flux were determined. The following decomposition of the 4-form flux $G$ is considered in \cite{AK}:
\begin{eqnarray}
G = & G_0+G_1 \\ \nonumber
G_0 \in & H^{1,1}(S_1, \R)\otimes H^{1,1}(S_2, \R) \\ \nonumber
G_1 \in & H^{2,0}(S_1, \C)\otimes H^{0,2}(S_2, \C) + h.c.
\end{eqnarray}
The conditions imposed in \cite{AK} are as follows:
\begin{equation}
\label{condition 3-brane in 5.1}
N_3 =0,
\end{equation}
and 
\begin{equation}
\label{decomp condition in 5.1}
G_0 =0.
\end{equation}
\par The condition (\ref{condition 3-brane in 5.1}) was relaxed in \cite{BKW} as follows:
\begin{equation}
N_3 \ge 0,
\end{equation}
and the extended list of the pairs of attractive K3 surfaces for which the tadpole cancels was obtained in \cite{BKW}, assuming the condition (\ref{decomp condition in 5.1}). 
\par The numbers of the pairs of the attractive K3 surfaces for which the tadpole is cancelled obtained in \cite{AK, BKW} are finite. Therefore, the complex structure moduli is stabilized for the situations discussed in \cite{AK, BKW}.
\par With respect to the Kummer surfaces $S_{[4 \hspace{1mm} 2 \hspace{1mm} 4]}$ and $S_{[4 \hspace{1mm} 0 \hspace{1mm} 4]}$ that are discussed in section \ref{ssec5.2}, the tadpole can be cancelled with flux when they are paired with appropriate attractive K3 surfaces. This is confirmed in section \ref{ssec5.2}. F-theory on K3 $\times$ K3 without insertion of flux yields four-dimensional theory with $N=2$ supersymmetry, and only adjoint representations of the gauge groups arise as matter. Because half of the supersymmetry is broken in the presence of flux, F-theory compactification on K3 $\times$ K3 with flux yields four-dimensional theory with $N=1$ supersymmetry. Hypermultiplet in four-dimensional theory with $N=2$ supersymmetry splits into a vector-like pair with flux; a vector-like pair arises in this situation when the tadpole cancels \cite{K}. In section \ref{ssec5.2}, we find that Kummer surfaces $S_{[4 \hspace{1mm} 2 \hspace{1mm} 4]}$ and $S_{[4 \hspace{1mm} 0 \hspace{1mm} 4]}$ belong to moduli of K3 surfaces constructed in section \ref{ssec4.2}, and we determine the exact matter spectra arising in F-theory on these Kummer surfaces times appropriate attractive K3 surfaces.

\subsection{Kummer surfaces belonging to the constructed moduli and vector-like pairs}
\label{ssec5.2}
\par F-theory compactification on K3 $\times$ K3 without flux yields four-dimensional $N=2$ theory, and the only light matter arising on the 7-branes in this compactification are adjoints of the gauge groups. Vector-like pairs can also arise when flux is turned on although it is necessary to confirm that tadpole cancels to ensure that they do not vanish owing to the anomaly. We observe that moduli of elliptic K3 surfaces constructed in section \ref{ssec4.2} include Kummer surfaces of specific complex structures. The results indicate that these Kummer surfaces are attractive K3 surfaces, and the tadpole can be cancelled when pairs of these Kummer surfaces times appropriately selected attractive K3 surfaces are considered. Thus, vector-like pairs arise in F-theory on these spaces.
\par The types of the elliptic fibrations \footnote{An elliptic K3 surface generally exhibits several distinct elliptic fibrations. See \cite{BKW} for a discussion of this in the context of F-theory compactification.} of attractive K3 surfaces, $S_{[4 \hspace{1mm} 2 \hspace{1mm} 4]}$ and $S_{[4 \hspace{1mm} 0 \hspace{1mm} 4]}$, are classified in \cite{Nish}. They are both Kummer surfaces \cite{PS-S}. The fibration types of the surface $S_{[4 \hspace{1mm} 2 \hspace{1mm} 4]}$ include fibration with Mordell--Weil rank 3 wherein the singularity type is $A_{11}D_4$. (Table 1.3, fibration no. 24 in \cite{Nish}) As reviewed in section \ref{ssec2.1}, the singular fibers of K3 surface obtained as a result of QBC are twice those of the original rational elliptic surface for general values of the parameters. Singular fibers collide when the parameters assume special values, and they are enhanced to another fiber type \cite{KRES}. Elliptic K3 surface (\ref{K3 rk 3 A5 in 4.2.6}) obtained in section \ref{sssec4.2.6} generally exhibits $A_5^2$ singularity with Mordell--Weil rank 3. When the parameters of QBC take special values, then two $I_6$ fibers collide, and they are enhanced to a type $I_{12}$ fiber. Thus, the singularity type $A_5^2$ enhances to $A_{11}$. The corresponding QBC is as follows: 
\begin{equation}
t \rightarrow \alpha_1 \, t^2.
\end{equation}
($\alpha_2=\alpha_3=0$ in (\ref{QBC in 2.1}).)
\par When the remaining parameters also assume special values, $A_{11}$ singularity is further enhanced in the moduli (\ref{K3 rk 3 A5 in 4.2.6}) to $A_{11}D_4$ singularity. The fibration with $A_{11}D_4$ singularity of Kummer surface $S_{[4 \hspace{1mm} 2 \hspace{1mm} 4]}$ corresponds to one such point in the moduli (\ref{K3 rk 3 A5 in 4.2.6}). Specifically, this indicates that the Kummer surface $S_{[4 \hspace{1mm} 2 \hspace{1mm} 4]}$ belongs to the moduli (\ref{K3 rk 3 A5 in 4.2.6}) constructed in section \ref{sssec4.2.6}. 
\par The tadpole can be cancelled in F-theory compactification with flux, when the Kummer surface $S_{[4 \hspace{1mm} 2 \hspace{1mm} 4]}$ is paired with the attractive K3 surface $S_{[2 \hspace{1mm} 1 \hspace{1mm} 2]}$ \cite{AK} by including sufficiently many 3-branes. Therefore, vector-like pairs arise from local rank-one enhancements of the singularities $A_{11}$ and $D_4$ on F-theory on the product $S_{[4 \hspace{1mm} 2 \hspace{1mm} 4]} \times S_{[2 \hspace{1mm} 1 \hspace{1mm} 2]}$. We consider the following local rank-one enhancements:
\begin{eqnarray}
A_{10} \subset A_{11} \\ \nonumber
A_3 \subset D_4.
\end{eqnarray}
Then the adjoints ${\bf 143}$ of $A_{11}$ and ${\bf 28}$ of $D_4$ decompose into irreducible representations of $A_{10}$ and $A_3$, respectively, as follows \cite{Sla}:
\begin{eqnarray}
{\bf 143} = & {\bf 120} + {\bf 11} (\, \ytableausetup{boxsize=.6em}\ytableausetup
{aligntableaux=center}\begin{ytableau}
 \
\end{ytableau} \,) +\overline{\bf 11} + {\bf 1} \\ \nonumber
{\bf 28} = & {\bf 15} + {\bf 6} (\, \ytableausetup{boxsize=.6em}\ytableausetup
{aligntableaux=center}\begin{ytableau}
 \\
 \\
\end{ytableau} \,) + \overline{\bf 6} +{\bf 1}.
\end{eqnarray}
Decomposition of ${\bf 28}$ of $D_4$ into irreducible representations of $A_3$ was discussed in \cite{MTmatter, K2}. Thus, the adjoints ${\bf 120}$ of $A_{10}$ and ${\bf 15}$ of $A_3$ arise on the 7-branes without fluxes. The vector-like pair ${\bf 11} +\overline{\bf 11}$ arises from $A_{11}$ singularity, and the vector-like pair ${\bf 6} + \overline{\bf 6} $ arises from $D_4$ singularity by including a flux. 
\par As previously seen in section \ref{sssec4.2.6}, a general member in the moduli (\ref{K3 rk 3 A5 in 4.2.6}) has  Mordell--Weil group isomorphic to $\Z^3$. When the singularity type enhances to $A_{11}D_4$ and K3 surface becomes the Kummer surface  $S_{[4 \hspace{1mm} 2 \hspace{1mm} 4]}$, then the Mordell--Weil group acquires the torsion part, and it becomes isomorphic to the following \cite{Nish}: 
\begin{equation}
\Z^3\times \Z_2.
\end{equation}
The global structure \cite{AspinwallGross, AMrational, MMTW} of the gauge group that forms on F-theory on $S_{[4 \hspace{1mm} 2 \hspace{1mm} 4]} \times S_{[2 \hspace{1mm} 1 \hspace{1mm} 2]}$ is:
\begin{equation}
SU(12) \times SO(8) / \Z_2 \times U(1)^3.
\end{equation}

\vspace{5mm}

\par Similarly, we find that the Kummer surface $S_{[4 \hspace{1mm} 0 \hspace{1mm} 4]}$ also belongs to the moduli (\ref{K3 rk 3 A5 in 4.2.6}). Kummer surface $S_{[4 \hspace{1mm} 0 \hspace{1mm} 4]}$ has a fibration of the Mordell--Weil rank 3 with the $A_{15}$ singularity. (Table 1.4, fibration no. 30 in \cite{Nish}) This surface corresponds to a point in the moduli (\ref{K3 rk 3 A5 in 4.2.6}), at which the singularity type is enhanced from the enhanced type $A_{11}$, that we discussed previously, further to $A_{15}$. From this, we find that the moduli (\ref{K3 rk 3 A5 in 4.2.6}) includes the Kummer surface $S_{[4 \hspace{1mm} 0 \hspace{1mm} 4]}$.
\par The tadpole can be cancelled in F-theory with flux, when the Kummer surface $S_{[4 \hspace{1mm} 0 \hspace{1mm} 4]}$ is paired with the Kummer surface $S_{[4 \hspace{1mm} 0 \hspace{1mm} 4]}$ \cite{BKW} \footnote{[2 0 2] in the list of \cite{BKW} represents the Kummer surface, which we denote by $S_{[4 \hspace{1mm} 0 \hspace{1mm} 4]}$ in this study.}. Under local rank-one enhancement:
\begin{equation}
A_{14} \subset A_{15},
\end{equation}
${\bf 255}$ of $A_{15}$ decomposes into the irreducible representations of $A_{14}$:
\begin{equation}
{\bf 255} = {\bf 224} + {\bf 15} (\, \ytableausetup{boxsize=.6em}\ytableausetup
{aligntableaux=center}\begin{ytableau}
 \
\end{ytableau} \,) + \overline{\bf 15} +{\bf 1}.
\end{equation}
The adjoints ${\bf 224}$ of $A_{14}$ arise on the 7-branes without fluxes. The vector-like pair ${\bf 15} + \overline{\bf 15}$ arises from the $A_{15}$ singularity in F-theory on $S_{[4 \hspace{1mm} 0 \hspace{1mm} 4]} \times S_{[4 \hspace{1mm} 0 \hspace{1mm} 4]}$ with a flux.
\par The Mordell--Weil group of Kummer surface $S_{[4 \hspace{1mm} 0 \hspace{1mm} 4]}$ with fibration of Mordell--Weil rank 3 with singularity type $A_{15}$ has the torsion part, and it is isomorphic to \cite{Nish}
\begin{equation}
\Z^3\times \Z_2.
\end{equation}
The following gauge group forms on F-theory on $S_{[4 \hspace{1mm} 0 \hspace{1mm} 4]} \times S_{[4 \hspace{1mm} 0 \hspace{1mm} 4]}$:
\begin{equation}
SU(16) / \Z_2 \times U(1)^3.
\end{equation}

\vspace{5mm}

\par Moduli of elliptically fibered K3 surfaces with the Mordell--Weil rank 4, the singularity type of which includes $A_4^2$, is constructed in \cite{Kimura1802}. Kummer surface $S_{[4 \hspace{1mm} 0 \hspace{1mm} 4]}$ has a fibration of Mordell--Weil rank 4, with the $A_7^2$ singularity, according to \cite{Nish}. (Table 1.4, fibration no. 52 in \cite{Nish}) This fibration of the surface $S_{[4 \hspace{1mm} 0 \hspace{1mm} 4]}$ corresponds to a point in the moduli of Mordell--Weil rank 4 built in \cite{Kimura1802}, at which $A_4^2$ singularity is enhanced to $A_7^2$. 
\par As stated previously, the tadpole can be cancelled in F-theory compactification with flux turned on, when the Kummer surface $S_{[4 \hspace{1mm} 0 \hspace{1mm} 4]}$ is paired with Kummer surface $S_{[4 \hspace{1mm} 0 \hspace{1mm} 4]}$. 
\par Under the following local rank-one enhancement
\begin{equation}
A_6 \subset A_7,
\end{equation}
${\bf 63}$ of $A_7$ decomposes into the irreducible representations of $A_6$ as:
\begin{equation}
{\bf 63} = {\bf 48} + {\bf 7} (\, \ytableausetup{boxsize=.6em}\ytableausetup
{aligntableaux=center}\begin{ytableau}
 \
\end{ytableau} \,) + \overline{\bf 7} + {\bf 1}.
\end{equation}
The adjoints ${\bf 48}$ of $A_6$ arise from the $A_7$ singularity without fluxes. The vector-like pair ${\bf 7} + \overline{\bf 7}$ also arises from the $A_7$ singularity in F-theory on $S_{[4 \hspace{1mm} 0 \hspace{1mm} 4]} \times S_{[4 \hspace{1mm} 0 \hspace{1mm} 4]}$ with flux.
\par The Mordell--Weil group of Kummer surface $S_{[4 \hspace{1mm} 0 \hspace{1mm} 4]}$ with fibration of Mordell--Weil rank 4 with singularity type $A_7^2$ has the torsion part, and it is isomorphic to \cite{Nish}
\begin{equation}
\Z^4\times \Z_2.
\end{equation}
The following gauge group forms on F-theory on $S_{[4 \hspace{1mm} 0 \hspace{1mm} 4]} \times S_{[4 \hspace{1mm} 0 \hspace{1mm} 4]}$:
\begin{equation}
SU(8)^2 / \Z_2 \times  U(1)^4.
\end{equation}

\section{Conclusions}
\label{sec6}
In this study, we constructed families of elliptically fibered K3 surfaces with the Mordell--Weil ranks 3 to 8 by utilizing QBC to glue pairs of identical rational elliptic surfaces. We considered F-theory compactifications on the resulting K3 surfaces times a K3 surface to yield models with 3 to 8 U(1) factors. These models yield four-dimensional theories with $N=2$ supersymmetry, and the enhanced supersymmetry imposes a strong constraint on the theories. We confirmed that the obtained theories satisfy the anomaly cancellation condition. F-theory models with multiple U(1) factors were investigated in recent studies. The explicit Weierstrass equations in this study added examples of such models. 
\par We also showed that Kummer surfaces of specific complex structures belonged to the moduli of K3 surfaces that were constructed in this study. We confirmed that the tadpole can be cancelled in F-theory compactification on these Kummer surfaces times appropriate attractive K3 surfaces with the flux turned on. We determined the exact matter spectra for these cases. 
\par Fibering K3 surfaces over the base $\P^1$ and over base surfaces to promote the geometries discussed in this study to elliptically fibered Calabi--Yau 3-folds and 4-folds and determine the Mordell--Weil groups of these spaces can be likely directions of future studies. It is expected that F-theory on the resulting spaces will yield 6d and 4d $N=1$ theories with multiple U(1) factors \footnote{4d $N=1$ F-theory models on elliptic Calabi-Yau 4-folds (not as direct products of K3 surfaces) without a U(1) gauge group are obtained in \cite{KCY4, Kdisc}. 6d $N=1$ F-theory models on elliptic Calabi-Yau 3-folds without a U(1) symmetry are discussed in \cite{Kimura1810, Kimura1902}.}. Additionally, Mordell--Weil ranks in the moduli spaces of elliptic K3 surfaces that were constructed in this study are generically constant although they enhance for special values of the parameters. (See \cite{KimuraMizoguchi} for a discussion.) This observation can aid in constructing models with higher Mordell--Weil groups.

\section*{Acknowledgments}

We would like to thank Shun'ya Mizoguchi and Shigeru Mukai for discussions. This work is partially supported by Grant-in-Aid for Scientific Research {\#}16K05337 from the Ministry of Education, Culture, Sports, Science and Technology of Japan.

\end{document}